\documentclass[pra,showpacs,notitlepage,10pt]{revtex4-1}
\usepackage{graphicx}
\usepackage{color}
\usepackage[T1]{fontenc}
\usepackage[utf8]{inputenc}
\usepackage{amssymb}
\usepackage{amsthm}
\usepackage{bbm}
\usepackage{mathtools}
\usepackage{float}

\def\tr{\operatorname{tr}}
\def\idty{{\leavevmode\rm 1\mkern -5.4mu I}} %  unit operator

\def\Rl{{\mathbb R}}\def\Cx{{\mathbb C}}
\def\Ir{{\mathbb Z}}

\def\norm #1{\Vert #1\Vert}

\def\braket#1#2{\langle #1,#2\rangle}

\def\bra#1{{\langle#1\vert}}
\def\ket #1{\vert#1\rangle}
\def\ketbra #1#2{{\vert#1\rangle\langle#2\vert}}
\def\kettbra#1{\ketbra{#1}{#1}}

\def\tr{\mathop{\rm tr}\nolimits}
\def\abs#1{\vert#1\vert}

\def\inv{^{-1}}

\def\re{\Re e}

\def\BB{{\mathcal B}}\def\HH{{\mathcal H}}\def\LL{{\mathcal L}}

\def\Xh{\widehat X}
\def\tor{{\mathbb T}}
\def\Fou{{\mathcal F}}
\def\cc#1#2{\raisebox{3pt}{$\scriptstyle\lfloor$}\raisebox{4pt}{$\scriptstyle#1$}\vert\raisebox{4pt}{$\scriptstyle#2$}\raisebox{3pt}{$\scriptstyle\rfloor$}}
\def\ccc#1#2{\overline{\cc{#1}{#2}}}
\def\pari{\Pi}

\def\intd#1{{\textstyle\int\!d#1\,}}

\def\UR{{\sf PUR}}
\def\MUR{{\sf MUR}}

\def\besselZ#1{{\mathrm z}_{#1}}
\def\Vh{\widehat V}

\begin{document}
\title{Uncertainty Relations for General Phase Spaces}

\author{Reinhard F. Werner}
	\email{reinhard.werner@itp.uni-hannover.de}
\affiliation{Institut f\"ur Theoretische Physik, Leibniz Universit\"at, Hannover, Germany}

\begin{abstract}
We describe a setup for obtaining uncertainty relations for arbitrary pairs of observables related by Fourier transform.  The physical examples discussed here are standard position and momentum, number and angle, finite qudit systems, and strings of qubits for quantum information applications. The uncertainty relations allow an arbitrary choice of metric for the distance of outcomes, and the choice of an exponent distinguishing e.g., absolute or root mean square deviations. The emphasis of the article is on developing a unified treatment, in which one observable takes values in an arbitrary locally compact abelian group and the other in the dual group. In all cases the phase space symmetry implies the equality of measurement uncertainty bounds and preparation uncertainty bounds, and there is a straightforward method for determining the optimal bounds.
\end{abstract}

\maketitle
%%%%%%%%%%%%%%%%%%%%%%%%%%%%%%%%%%%%%%%%%%%%%%%%%%%%%
\section{Introduction}
Uncertainty relations are quantitative expressions of two fundamental features of quantum mechanics. The first feature is the observation that there are {\it no dispersion free states} \cite[Sect.~IV.1]{vonNeumann}. That is, we cannot find states, which give fixed (non-statistical) results on all observables. This  is already seen for many pairs of observables, most famously for position and momentum. The well-known Heisenberg-Kennard relation is a ``preparation uncertainty relation'', i.e., a quantitative expression of the observation that there is no quantum state for which both the position distribution and the momentum distribution are sharp. The second feature is loosely referred to as {\it complementarity}, or the existence of mutually exclusive experimental arrangements. More precisely, there are measurements which cannot be simulated as marginals of a joint measurement device. The word ``simulated'' here indicates that complementarity runs much deeper than the trivial impossibility to build two experiments on top of each other. The basic impossibility statement is in terms of observables (positive operator valued measures), which encode just the statistical ``quantum input to classical output'' behaviour of a device. Two measurement devices may thus be incompatible in the sense that it is impossible to build a new device with two kinds of outcomes, such that ignoring any one of them leaves one with a device statistically equivalent to one of the given ones. There is a particular way of attempting such a joint measurement, namely by measuring first one observable, and then trying to retrieve the other from the post-measurement state. For complementary observables this is bound to fail, which is another way of saying that the first measurement necessarily disturbs the system. A ``measurement uncertainty relation'' is a quantitative expression of the non-existence of joint measurements, and hence also encodes the {\it error-disturbance tradeoff} associated with measurements.

The preparation and the measurement aspect of uncertainty are logically independent since they refer to quite different experimental scenarios. An experiment verifying a preparation uncertainty relation between observables $A$ and $B$ will separately determine the distributions of $A$ and $B$, so that no individual particle is subject to both kinds of measurement. The minimum uncertainty objects in this case are states. In contrast, for measurement uncertainty an $A$-value and a $B$-value is obtained for each particle, often in succession as in the error-disturbance scenario. The minimum uncertainty objects are approximate joint measurements of $A$ and $B$. There is no direct operational link between these scenarios, and the quantitative bounds for preparation and measurement uncertainty are, in general, different. Indeed for two projection valued (standard) observables, which are for this purpose mainly specified by their eigenbases, preparation uncertainty is zero if the two bases share {\it one} eigenvector, whereas measurement uncertainty vanishes when the observables commute, i.e., when they share {\it all} eigenvectors. The distinction is also borne out by the detailed study of angular momentum uncertainty \cite{URAM}, where preparation and measurement uncertainty require quite different methods. It is therefore somewhat surprising that for the case of position and momentum \cite{Wer04,BLW1,BLW2} the measurement uncertainty relations are quantitatively the same as the preparation uncertainty relations. The abstract reason for this is {\it phase space symmetry}.

However, this type of symmetry and the result mentioned is by no means restricted to the standard position/momentum case. The purpose of this paper is to review the application of these ideas to other phase spaces. The common features of the systems considered are the following: One has a pair of observables, which we will just continue to call position and momentum, which are related to each other by Fourier transform. Position will take values in a space $X$ (generalizing $X=\Rl^n$), on which translations make sense, so we take it as an abelian group. The unitary operators implementing translations in position space will be functions of momentum. Symmetrically, there is a momentum space $\Xh$ whose translations are generated by unitaries which are functions  of position. Such pairs appear in many traditional systems in physics, e.g., number/phase, or lattice site/quasi-momentum. Quantum information has additionally generated a lot of interest in finite cases, like qudit systems or qubit strings. For qubit strings the position observable is the readout of strings in $Z$ basis and momentum the readout in $X$ basis. A typical uncertainty question here would be how accurately an eavesdropper can possibly measure a string in one basis without disturbing the readouts in another basis, when both errors are ascertained, for example, in Hamming distance.

Since we claim the quantitative agreement of measurement and preparation uncertainty bounds, we need to express the bounds by a uniform set of criteria. It turns out that all it takes is to fix, for each observable, a metric on the outcome space, together with a certain exponent. This allows closely connected definitions of variances for preparation uncertainties and the distance of probability distributions needed for measurement uncertainty. To summarize, each case of the theory developed here involves the following choices:
\begin{itemize}
\item a {\it phase space} $\Xi=X\times\Xh$, which is given in terms of a locally compact abelian group $X$ and its dual $\Xh$. We will refer to $X$ as the position space, and to $\Xh$ as the momentum space.
\item a translation invariant {\it metric} on the space $X$, and another one on $\Xh$.
\item {\it error exponents} $1\leq\alpha,\beta\leq\infty$, which determine whether the error measures gives more emphasis to small or to large distances.
\end{itemize}
We will develop the theory in full generality, for any such choice. This includes the equality of measurement and preparation uncertainty bounds. The bounds are best expressed in terms of the set of achievable pairs $(\Delta P,\Delta Q)$ of uncertainties, and especially the trade-off curve of pairs where neither uncertainty can be reduced without increasing the other. There is a concrete prescription how to calculate this curve: Each point on the curve is determined by finding the ground state of a certain operator, and this solution also gives a corresponding minimum uncertainty state (resp.\ minimum uncertainty joint measurement). Sometimes the ground state problem is very simple. For example, the standard position/momentum case leads to the problem of finding the ground state of a harmonic oscillator, making pure Gaussians the minimum uncertainty states. This case also has an additional dilatation symmetry, so that with each achievable uncertainty pair also the hyperbola $(\lambda\Delta P,\lambda^{-1}\Delta Q)$ is achievable. Therefore, the uncertainty region is completely described by the lowest lying hyperbola, i.e., by the lowest product $\Delta P\,\Delta Q$. However, this is the only case in which an uncertainty product adequately describes the trade-off curve. The general ground state problem cannot be solved in such a simple form. Therefore we look at concrete cases, listed in Table~\ref{tab:choose}, selected in part for their physical interest and in part to illustrate some of the features that may occur.

\begin{table}\begin{centering}
\begin{tabular}{r||c|c|c|c|c|c}
  System          & $X$   & $\Xh$ & metrics     & $\alpha,\beta$& Section& Ref\\ \hline
  Canonical pair  &$\Rl$  & $\Rl$ & abs         & all           &&  \cite{BLW2}  \\
  Mechanics       &$\Rl^n$& $\Rl^n$ & Euclidean & 1,2           &\ref{sec:exStandard}            \\ \hline
  angle/number    & $\tor$&       & arc or chordal  & 2         &&  \cite{KW} \\
                          &&$\Ir$   &discrete or abs &  1,2     &\ref{sec:exNumber}&\\    \hline
  qudit           &$\Ir_d$&$\Ir_d$& discrete     & 1            &\ref{sec:exQudit}&\\
  qubit string    &$\Ir_2^N$&$\Ir_2^N$& Hamming& all            &\ref{sec:exStrings}&
\end{tabular}
 \caption{\label{tab:choose}Phase spaces and parameters considered in this paper}
\end{centering}\end{table}

The paper is organized as follows: We will review the basic theory of phase space quantum mechanics in the next section, and the relevant notions of uncertainty in Sect.~\ref{sec:unc}. This is followed in Sect.\ref{sec:examp} by discussing the special instances, as summarized in Table~\ref{tab:choose}.

\section{Phase spaces}\label{sec:ps}
In this chapter we outline phase space quantum mechanics in the general setting outlined above. The generality forces us to use a relatively abstract (i.e., mathematical) language. Physicists feeling not so comfortable with this level of abstraction should read this section and the next with two concrete examples from Table~\ref{tab:choose} in mind, one of which should be the ``standard'' case of one position/momentum pair.

The origin of the theory outlined here is in \cite{QHA,QHAcorr}, where it is carried out for the standard phase spaces $\Rl^n\times\Rl^n$. The generalization to general phase spaces is straightforward for the parts we need for the current context, and only needs some standard results of the harmonic analysis (Fourier theory) of locally compact abelian groups \cite{Reiter,Hewitt}. A detailed treatment, also of the fine points, is in preparation with Jussi Schultz.

We assume from on now that a group $X$ of ``position shifts'' is given. Technically, any locally compact abelian group is allowed, but in physical or QI applications we will be talking about one of the groups from Table~1. Apart from one or more canonical degrees of freedom, like position/momentum of quantum optical field quadratures, we may also have angle or phase variables with an intrinsic periodicity given by the group $\tor$ of phases (complex numbers with modulus $1$ under multiplication) or, equivalently $\tor=\Rl/(2\pi\Ir)$. Further there may be discrete variables given by integers and either unbounded ($X=\Ir$) or modulo some number $d$ ($X=\Ir_d$). Furthermore, arbitrary combinations of these choices are allowed.

We will denote integration with respect to the Haar measure on $X$ by ``$\int\!dx$''. This measure is unique up to a constant, and is characterized by its translation invariance, i.e., by the possibility to substitute a shifted variable without functional determinant factors. In the discrete cases it is often natural to give each point unit measure. Integrals with respect to this ``counting measure'' are just sums over $x$. In the compact cases ($\tor$ and $\Ir_d$ and their products) the total Haar measure is finite, and it is often convenient to take it as a probability measure, i.e., normalized to $1$. Note that these natural choices are in conflict for $\Ir_d$, which means that we have to make choices very much analogous to where one wants to stick the normalization factors $2\pi$ for the standard Fourier transform.

The basic Hilbert space of our systems will now be $\HH=\LL^2(X,dx)$, the square integrable functions on $X$. In it the projection valued position observable acts by multiplication operators, i.e., the position probability density associated with a vector $\psi$ will be $\abs{\psi(x)}^2$.  The unitary shift operators $(U_x\psi)(y)=\psi(y-x)$ are clearly not functions of position. But since the underlying group $X$ is abelian, they commute and can therefore by jointly diagonalized, i.e., be represented as multiplication operators in another representation. This will, of course, be the momentum representation reached by the Fourier transform. The Fourier transform of a function $\psi:X\to\Cx$ will be a function $\Fou\psi:\Xh\to\Cx$, where $\Xh$ is the dual group of $X$. This is abstractly the set of continuous multiplicative functions from $X$ to $\tor$. If $p\in\Xh$ labels such a function, we write it as $x\mapsto\cc px$. By definition,  $\cc p{x_1+x_2}=\cc p{x_1}\cc p{x_2}$. The sum in $\Xh$ is defined by $\cc{p_1+p_2}x=\cc{p_1}x\cc{p_2}x$. Concretely, when
$X=\Rl^n$ we also have $\Xh=\Rl^n$ and $\cc px=\exp(ip\cdot x)$, where the dot denotes the scalar product. Similarly, for the pair $\Xi=\tor\times\Ir$ we have
$\cc\alpha n=\exp(i\alpha n)$. Note that changing $\alpha$ here to $\alpha+2\pi$ (which represents the same element in $\tor$) does not change the value of $\cc\alpha n$, and this property is what forces $n\in\Ir$. The same reasoning leads to the form $\cc px=\exp(2\pi i px/d)$ for $\Xi=\Ir_d\times\Ir_d$. Now the Fourier transform and its inverse are defined by
\begin{equation}\label{fou}
   (\Fou\psi)(p)=\int\!dx\, \ccc px\,\psi(x)   \qquad\text{and}\qquad (\Fou^*\phi)(x)=\int\!dp\, \cc px\,\phi(x)
\end{equation}
Here the overbar means complex conjugation. Note that each of these formulas fixes a normalization of the Haar measure on $\Xh$ relative to that on $X$ and it is a Theorem that these two potentially distinct conventions do coincide \cite[Thm.4.4.14]{Reiter}.  $\Fou$ is unitary operator with inverse $\Fou^*$, and the usual formulas relating the product of functions to the convolution of their Fourier transforms hold, with the pertinent powers of $2\pi$ absorbed into the definition of the measures. The momentum observable acts by multiplication after Fourier transform, and the momentum probability density associated with a state vector $\psi$ is just $\abs{(\Fou\psi)(p)}^2$.

As an example let us consider a qubit, which is usually not looked at in these terms. The group here is $X=\{0,1\}$ a single bit with addition mod 2. The dual group $\Xh$ is the same with $\cc px=\exp(2\pi i px/2)=(-1)^{px}$. The Fourier transform acts on $\HH=\Cx^2$ by a Hadamard matrix. The position observable is given by the diagonal matrices (functions of $\sigma_z$ and the momentum observable is given by the functions of $\sigma_x$.

Momentum translations will act by in the position representation multiplication by multiplication with $\cc px$. Combining these with position translations we get the phase space translation operators, or {\it Weyl operators}
\begin{equation}\label{weyl}
  (W(q,p)\psi)(x)=\cc px\psi(x+q).
\end{equation}
These form a projective representation of the phase space translation group $\Xi=X\times\Xh$:
\begin{equation}\label{weyladd}
  W(q_1,p_1)W(q_2,p_2)=\cc{p_1}{q_2}W(q_1+q_2,p_1+p_2)
\end{equation}
Sometimes it is customary to change each Weyl operator by a phase, particularly in the standard $\Rl^n\times\Rl^n$ case, where this simplifies the relation for the adjoint to $W(\xi)^*=W(-\xi)$. With the choice \eqref{weyl} this reads instead
\begin{equation}\label{weyladj}
  W(q,p)^*= \cc pq\, W(-q,-p).
\end{equation}
The factor in \eqref{weyladd} depends on phase conventions, but the commutation phase
\begin{equation}\label{weylcom}
  W(q_1,p_1)W(q_2,p_2)= \cc{p_1}{q_2}\ccc{p_2}{q_1} W(q_2,p_2)W(q_1,p_1)
\end{equation}
does not.

For many purposes it is not necessary for the notation to separately refer to position and momentum, so we will just write $\xi\in\Xi$ for the pair $\xi=(q,p)\in X\times\Xh$ and ``$d\xi$'' for ``$dq\,dp$''.  With the above conventions about normalizing the measures, this translates for standard phase space into $dp\,dq/(2\pi)$. Thus phase space volume is measured in units of Planck's constant $h=2\pi\hbar=2\pi$. It should be noted that while the normalizations of the individual measures $dq$ and $dp$ contain a conventional factor, the phase space measure is independent of such conventions. The phase space translations of quantum observables (operators $A\in\BB(\HH)$) and classical observables (functions $f:\Xi\to\Cx$) are now given by
\begin{equation}\label{alpha}
  \alpha_\xi(A)= W(\xi)^*AW(\xi)   \qquad\text{and}\qquad
  (\alpha_\xi f)(\eta)=f(\eta-\xi)
\end{equation}
Similarly, we can define the operation of phase space inversion by the parity operator $(\pari\psi)(x)=\psi(-x)$ as
\begin{equation}\label{betaparity}
  \beta_{-}(A)= \pari A\pari   \qquad\text{and}\qquad
  (\beta_{-} f)(\eta)=f(-\eta).
\end{equation}
This notation is chosen to emphasize the quantum-classical analogy, and helps to generalize the {\it convolution} from phase space functions to operators\cite{QHA}. Indeed, the convolution of functions can be alternatively written as
\begin{equation}\label{convolveff}
\begin{array}{rl}
  (f\ast g)(\xi)&=\int\!d\eta\, f(\eta) g(\xi-\eta)\\
                &=\int\!d\eta\, f(\eta) (\alpha_\eta g)(\xi)  %\\&
                =\int\!d\eta\, f(\eta) (\alpha_\xi \beta_{-}g)(\eta) \\
       (f\ast g)&=\int\!d\eta\, f(\eta) (\alpha_\eta g),\\
\end{array}
\end{equation}
where the last line is just a version of the second, read as an equation between functions. This version allows the definition of the convolutions between functions and operators (giving an operator), and the second expression in the second line, with the trace of functions substituted for phase space integrals, suggests the convolution of two operators, which is then again a phase space function:
\begin{eqnarray}\label{convolvefA}
  f\ast A &=& A\ast f =\int\!d\eta\, f(\eta) (\alpha_\eta A)\\
  (A\ast B)(\xi) &=& \tr\bigl(A (\alpha_\xi\beta_{-} B)\bigr) \label{convolveAB}
\end{eqnarray}
We have not specified the analytic conditions for these integrals to exist: Even just for functions on an infinite phase space (like $f=g=1$) the integral may diverge. A crucial Lemma in this theory, based on the square integrability of matrix elements $\braket\phi{W(\xi)\psi}$, is that if all factors involved are either integrable functions (i.e., $\norm f_1=\int\!d\xi\, \abs{f(\xi)}<\infty$) or  ``trace class'' operators (i.e., $\norm A_1=\tr\abs A<\infty$) then the same holds for their convolution. Convolution is then a commutative and associate product, and determines a Banach algebra with the $1$-norm. It also has the crucial property that the convolution of positive factors is positive. The convolution can also be extended to the case where one factor is just a bounded function or operator. However, in this case the result can only be guaranteed to be bounded, and in a product of several factors we can usually only allow one such factor.

The main upshot of this formalism for our purpose is the characterization of {\it covariant phase space observables}. By definition, these are normalized positive $\BB(\HH)$-valued measures $F$ that commute with phase space translations. We use the compact notation $F[f]=\int\!F(d\xi)\,f(\xi)$, i.e., $F$ with round parentheses is a function on subsets of $\Xi$, and $F$ with brackets is the linear operator $F[\cdot]:\LL^\infty(\Xi,d\xi)\to\BB(\HH)$ one gets from this by integration. Since, conversely, $F(\sigma)=F[\chi_\sigma]$, with $\chi_\sigma$ the indicator function of a measurable set $\sigma\subset\Xi$, we consider these two to be essentially the same object. Covariance then means that $\alpha_\xi F[f]=F[\alpha_\xi f]$. Then the basic theorem on the subject \cite[Prop.3.3]{QHA} states that the covariant phase space observables are in one-to-one correspondence with the density operators $\rho_F$ on $\HH$, given by the formula
\begin{equation}\label{covobs}
  F[f]=\rho_F\ast f.
\end{equation}
Calling $\rho_F$ here a density operator has a double meaning: On one hand, it describes the conditions for the correspondence $F\leftrightarrow\rho_F$, namely $\rho_F\geq0$ and $\tr\rho_F=1$. Somewhat accidentally, these are the conditions for an operator describing a mixed quantum state. On the other hand, the measure $F(\cdot)$ has an operator valued density with respect to $d\xi$, namely the translates $\alpha_\xi(\rho_F)$. This ``accident'' will be crucial later for establishing the equivalence between measurement uncertainty relations for $F$ and preparation uncertainty relations for a certain state, namely $\rho_F$.

For measurement uncertainty we need the position and momentum marginals of such observables, i.e., the expectations of functions of only position or only momentum. So let $f:X\to\Cx$ be some function on position space. We can consider it either as a classical function on phase space $f_q:\Xi\to\Cx$ by $f_q(x,p)=f(x)$, or as a quantum operator $f(Q)\in\BB(\HH)$, as determined by the functional calculus. This is the multiplication operator $(f(Q)\psi)(x)=f(x)\psi(x)$. Then expectations can alternatively be written as an integral over phase space (resp.~a trace) or as an integral over just $X$ with respect to a suitable ``marginal''. Thus if $\rho$ is a density operator and $\mu$ is a probability density on phase space, we define marginals $\mu^q$ and $\rho^Q$ by
\begin{eqnarray}\label{marginals}
  \intd x \mu^q(x) f(x)&=& \intd \xi \mu(\xi)f_q(\xi)\\
  \intd x \rho^Q(x) f(x)&=& \tr\rho f(Q)
\end{eqnarray}
Thus $\rho^Q$ is just the position probability density associated with the quantum state $\rho$. Classically, $\mu^q$ arises from $\mu$ by integrating out the momenta. Similarly, in the quantum case, integrating over all momentum translates of $\rho$ produces an operator, namely $\intd p\, \alpha_{(0,p)}(\rho)=(\rho^Q)(Q)$. Now suppose we have prepared a quantum state $\rho$, measure the covariant observable $F$, and evaluate the expectation of a function $f_q$  depending only on position. Then the overall expectation is
\begin{eqnarray}\label{expectQ}
  \tr\rho(\rho_F\ast f_q)&=&\rho\ast \beta(\rho_F\ast f_q)(0)=\int\!dx\ (\rho\ast\beta\rho_F)^q f(x)=\int\!dx\ \mu^q(x)f(x)\\
  \mbox{with}\quad
  \mu^q&=& \rho^Q\ast (\beta\rho_F)^Q
\end{eqnarray}
This has a remarkable interpretation, which is the basis of the equivalence between measurement and preparation uncertainty in our setting: The probability density for the position marginal of a covariant observable $F$ in the state $\rho$ is the convolution of the density $\rho^Q$ for the ideal position observable in the same state and the corresponding density of another state, $\beta\rho_F$. Since convolution is the operation representing the sum of independent random variables we arrive at the following statement:
\begin{quote}{\it The position marginal of a covariant phase space observable can be simulated by first making an ideal position measurement
  and adding to the outcome some random noise with a fixed distribution, independent of the input state. The distribution of the noise is the position of another quantum state characterising the observable.}
\end{quote}
Of course, the same holds mutatis mutandis for momentum (and letters $p,P$ replacing $q,Q$), with {\it the same} state $\beta\rho_F$ characterizing the observable. Thus the preparation uncertainty tradeoff of having either $(\beta\rho_F)^Q$ or $(\beta\rho_F)^P$ sharp translates directly into the measurement uncertainty tradeoff of measuring either position or momentum precisely, but never both.

\section{Measurement and preparation uncertainty}\label{sec:unc}
The statement that measurement and preparation uncertainty bounds are quantitatively equal for phase space observables presupposes that the errors and variances are defined in a closely related way. This begins by choosing, for each observable a metric $d$ on the set $X$ of outcomes. This not only fixes the units in which all deviations are measured, but also is an adaptation to the concrete problem at hand. For example, for discrete outcomes we might just be interested in whether outcomes coincide, without assigning a numerical weight (other than a constant) to their distance in case they don't. This is then simply expressed by choosing the discrete metric  $d(x,y)=1-\delta_{xy}$. For real valued observables like position and momentum we always take the standard distance $d(x,y)=\abs{x-y}$. The only requirement on the metric will be that it is translation invariant, i.e., $d(x+z,y+z)=d(x,y)$.

In addition we will fix, for every observable an error exponent $\alpha$ with $1\leq\alpha\leq\infty$. Then if $\mu$ is a probability measure on $X$, we define its {\it deviation from a point} $x\in X$ as
\begin{equation}\label{deviate}
  d(\mu,x)=\left(\int\mu(dy) d(x,y)^\alpha\right)^{\frac1\alpha}.
\end{equation}
So, for example, for $\alpha=2$ we get the mean quadratic deviation, for $\alpha=1$ the mean absolute deviation and in the limit $\alpha\to\infty$ the maximal deviation (discounting sets of $\mu$-measure zero). The {\it spread} of a probability measure, which we just denote by $d(\mu)$ is its smallest deviation from any point, i.e.,
\begin{equation}\label{spread}
  d(\mu)=\min_x d(\mu,x).
\end{equation}
The notation \eqref{deviate} suggests that this expression somehow extends the original metric on $X$ to one on the probability measures. This is intentional, and for the formulation of measurement uncertainty we actually also need the extension to the case where both arguments are probability measures, say $\nu$ and $\mu$. In this case we set
\begin{equation}\label{Wstein}
  d(\nu,\mu)=\inf_\gamma\left(\int\gamma(dx\,dy) d(x,y)^\alpha\right)^{\frac1\alpha},
\end{equation}
where the infimum is over all ``couplings'' of $\mu$ and $\nu$, i.e., all joint distributions on $X\times X$ such that the first variable is distributed according to $\nu$ and the second according to $\mu$. When $\nu$ is concentrated on the point $x$ this expression reduces to \eqref{deviate}. This metric is called the {\it transport metric} \cite{Villani} associated with $d$ and $\alpha$. It expresses the minimal cost of converting $\nu$ into $\mu$, when transferring one mass unit from $x$ to $y$ costs $d(x,y)^\alpha$. In particular, when $\mu=\nu$, the best coupling (=transport plan) is to leave everything as is, so corresponds to $\gamma$ spread out on the diagonal of $X\times X$, giving $d(\mu,\mu).=0$ Similarly, when $\mu$ arises from $\nu$ by translation of the variable by $a$, we have $d(\mu,\nu)=d(a,0)$. Finally, for a convolution of probability measures we get
\begin{equation}\label{dconvolve}
  d(\mu\ast\nu,\mu)\leq d(\nu,0).
\end{equation}

Using this notation we can say that preparation uncertainty theory for the observables $P$ and $Q$ is the study of the set of pairs
\begin{equation}\label{URregion}
  \UR=\Bigl\{\bigl(d(\rho^P),d(\rho^Q)\bigr)\Bigm| \rho\ \mbox{ a state}\Bigl\},
\end{equation}
where $\rho^P,\rho^Q$ denote the position and momentum distributions of the state $\rho$. In particular, we want to show that this ``uncertainty region'' contains no points near the origin.

Measurement uncertainty is a property of any (approximate) joint measurement $F$ of two observables. For each of them, i.e., in our case $P$ and $Q$, we compare the output marginal distributions in a state $\rho$, denoted by $\rho F^Q$ and $\rho F^P$, with what one would have got with the corresponding ideal measurement. We want the result to be uniformly good for all input states, i.e., we look at
\begin{equation}\label{mUR}
  d(F^Q,Q)=\sup_\rho d\bigl(\rho F^Q,\rho^Q\bigr),
\end{equation}
and the corresponding quantity for $P$. This vanishes if and only if the position distribution obtained by $F$ is the same as the usual one for arbitrary input states $\rho$. In that case, Heisenberg \cite{Heisenberg1927} told us to expect that the corresponding quantity fails badly for momentum. The tradeoff is thus given by the measurement uncertainty region
\begin{equation}\label{MURregion}
  \MUR=\Bigl\{\bigl(d(F^P,P),\,d(F^Q,Q)\bigr)\Bigm| F\ \mbox{ a joint measurement}\Bigl\}.
\end{equation}
For variants and a discussion of these notions, see \cite{BLW1}.

Now for a covariant measurement $F$ it is easy to compute both measurement uncertainties. Combining \eqref{expectQ} with \eqref{dconvolve} we get the bound in terms of the ``noise generated by $F$'', i.e.,
\begin{equation}\label{estimateConv}
  d\bigl(\rho F^Q,\rho^Q\bigr)\leq d((\beta\rho_F)^Q,0)\leq d\bigl((\beta\rho_F)^Q\bigr).
\end{equation}
Here the last inequality holds with equality iff the position distribution $(\beta\rho_F)^Q$ has mean zero, which can be achieved easily by just shifting all position outcomes. Any other choice of a constant offset would be clearly sub-optimal, so we have equality in the optimal case. Moreover, equality holds in \eqref{dconvolve} if $\mu$ is a point measure, so because of the supremum in \eqref{mUR}, we have
\begin{equation}\label{MUisPU}
  d(F^P,P)=d\bigl((\beta\rho_F)^P\bigr),\quad\mbox{and}\quad d(F^Q,Q)=d\bigl((\beta\rho_F)^Q\bigr)
\end{equation}
for all covariant (and centered) $F$. Hence measurement uncertainties for $F$ are the same as the preparation uncertainties for the state $\beta\rho_F$. That the general case can be reduced to the covariant one by an averaging procedure was shown in \cite{Wer04,BLW2}. Hence we have
\begin{equation}\label{MURisPUR}
  \MUR=\UR.
\end{equation}

\section{How to compute the bounds from a ground state problem}\label{sec:groundstate}
General methods for efficiently computing measurement uncertainty relations are still scarce. We therefore use the known methods for preparation uncertainty. The first observation is that it is better to work with variances than with deviations, i.e., to omit the roots in definition \eqref{deviate}. For the purposes drawing uncertainty diagrams this is just a rescaling, but the linearity in $\rho$ makes estimates more straightforward. The second observation is that we can reduce to the case of centered states, for which the minimum in \eqref{spread} is attained at $x=0$ (resp. $p=0$). This can always be achieved by a translation. Hence for the position variance we just have to compute the expectation of the function $x\mapsto d(x,0)^\alpha$, or, written in the functional calculus the expectation of the unbounded operator $d(Q,0)^\alpha$. The tradeoff is taken into account by considering linear combinations of variances with positive weights, and minimizing these over all states. That is, for $t>0$:
\begin{equation}\label{linbound}
  d(\rho^P)^\beta+t d(\rho^Q)^\alpha=\tr\rho \Bigl(d(P,0)^\beta+t\,d(Q,0)^\alpha\Bigr)\equiv \tr\rho H_{\alpha\beta}(t).
\end{equation}
It is clear that the operator $H_{\alpha\beta}(t)$ appearing here is usually unbounded, but positive, so technically speaking we mean its Friedrichs extension. In all cases it has discrete spectrum, and minimizing the above expression over $\rho$ is just finding its ground state ``energy'' $E_{\alpha\beta}(t)$. Essentially, this function is the Legendre transform of the tradeoff curve we want to to determine: For fixed value of $d(\rho^Q)$ we find the best otherwise state-independent bound on $d(\rho^P)$ by treating $t$ as a parameter to be optimized. This gives the state independent bound
\begin{equation}\label{toCurve}
  d(\rho^P)^\beta\geq \sup_t\Bigl\{E_{\alpha\beta}(t)-t\, \Delta\Bigr\}\quad\mbox{if}\ \Delta= d(\rho^Q)^\alpha.
\end{equation}
This is the description of the tradeoff curve (or rather: its best convex approximation), and the following examples will all be based on this method.

\section{Examples}\label{sec:examp}
Here we will provide the more concrete examples of the theory outlined in the previous sections.

\subsection{The standard case: $\Xi=\Rl^n\times\Rl^n$ with Euclidean distance }\label{sec:exStandard}
\begin{figure*}[t]
	\includegraphics[width=0.45\textwidth]{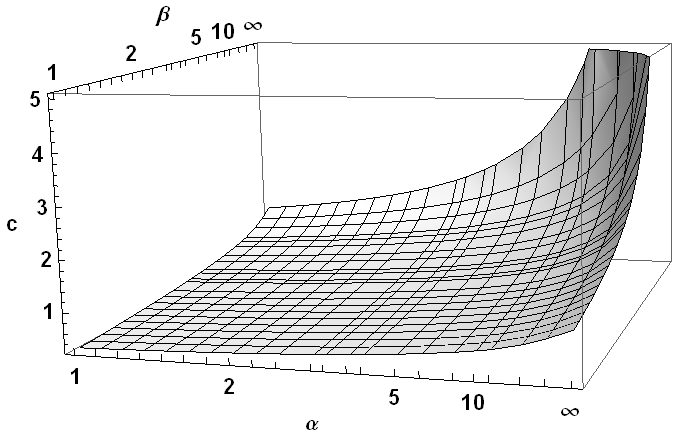}
	\caption{Uncertainty constants $c_{\alpha\beta}(1)$. Axes have been scaled non-linearly to represent the infinite range. We have $c_{\infty\infty}(1)=\infty$, because there are no states with strictly bounded support
     in both position and momentum.
	\label{fig:Cab}}
\end{figure*}

Due to the dilation symmetry $(x,p)\mapsto(\lambda x,\lambda\inv p)$ the uncertainty region will be bounded by a hyperbola, and completely described by the best constant $c$ in
\begin{equation}\label{std}
  d(\rho^Q)\,d(\rho^P)\geq c_{\alpha,\beta}(n)\hbar.
\end{equation}
This scaling symmetry is what makes ``dimensional analysis'' work, so in the above relation we brought in the dimensional constant $\hbar$ to make $c$ dimensionless, but will take $\hbar=1$ in the sequel. The textbook case is $c_{2,2}(1)=1/2$, For $n=1$ the constants (from \cite{BLW2}) are shown in Fig.~\ref{fig:Cab}

Depending on the application there may be good reason to explore other exponents than $2$. For example, $\alpha=\infty$ corresponds to the case of strict spatial confinement, like for the lateral position on passing a slit. If we are interested in the root mean square momentum spread after the slit, the constant $c_{\infty,2}$ will give a much better bound than first converting the slit information to a constraint on the root mean square deviation in position, and using $c_{2,2}$ instead.

In order to relate the constants $c_{\alpha,\beta}(n)$ to a ground state problem we consider the two-parameter family of Hamiltonians and ground states
\begin{equation}\label{Hab}
  H_{\alpha\beta}(a,b)=a\,\Bigl(\sum_{i=1}^n Q_i^2\Bigr)^{\textstyle\frac\alpha2}+b\,\Bigl(\sum_{i=1}^nP_i^2\Bigr)^{\textstyle\frac\beta2}
                 \geq E(a,b)\idty.
\end{equation}
Then $E$ satisfies the identities $E(\mu a,\mu b)=\mu E(a,b)$ from homogeneity, and $E(\lambda^\alpha a,\lambda^{-\beta}b)=E(a,b)$ from dilation symmetry so that
\begin{equation}\label{Eab}
  E(a,b)=a^{\textstyle\frac\beta{\alpha+\beta}}\ b^{\textstyle\frac\alpha{\alpha+\beta}}\ E,
\end{equation}
with $E=E(1,1)$ for short.
We now optimize $\lambda$ on the right hand side of the following inequality to get
\begin{equation}\label{cabfromEab}
  E\leq \lambda^\alpha d(\rho^Q)^\alpha+ \lambda^{-\beta}d(\rho^P)^\beta
         =(\alpha +\beta )\ \alpha ^{-\frac{\alpha }{\alpha +\beta }} \beta ^{-\frac{\beta}{\alpha +\beta }}\
           \Bigl(d(\rho^Q)d(\rho^P)\Bigr)^{\textstyle\frac{\alpha\beta}{\alpha+\beta}},
\end{equation}
which shows \eqref{std}.

The dimension dependence is straightforward in the quadratic case, since variances just add up to give Euclidean variance, i.e., $H_{\alpha\beta}$ separates in Cartesian coordinates. We get $E=n$, and hence
\begin{equation}\label{22multidim}
  c_{2,2}(n)=\frac n2.
\end{equation}
In general we can still use the rotation symmetry to simplify the problem, seeking joint eigenfunctions of  $H_{\alpha\beta}$, and the angular Laplacian $L^2$. If the eigenvalue for the latter operator is $\lambda$, we have to find the smallest $E$ for which we can solve the radial equation
\begin{equation}\label{radial}
  r^\alpha \phi(r)+\Bigl(-\frac{d^2}{dr^2}+ \frac{4\lambda+(n-1)(n-3)}{4r^2}\Bigr)^{\textstyle\frac\beta2}\phi(r)=E\,\phi(r).
\end{equation}
We have chosen here to include the weight coming from the integration in polar coordinates into the wave function, so the radial Laplacian, i.e., the operator in parentheses, contains no first derivatives. Of course, unless $\beta$ is even, this is not a differential operator of finite order. No general solution is available.

\begin{figure*}[ht]
	\includegraphics[width=0.45\textwidth]{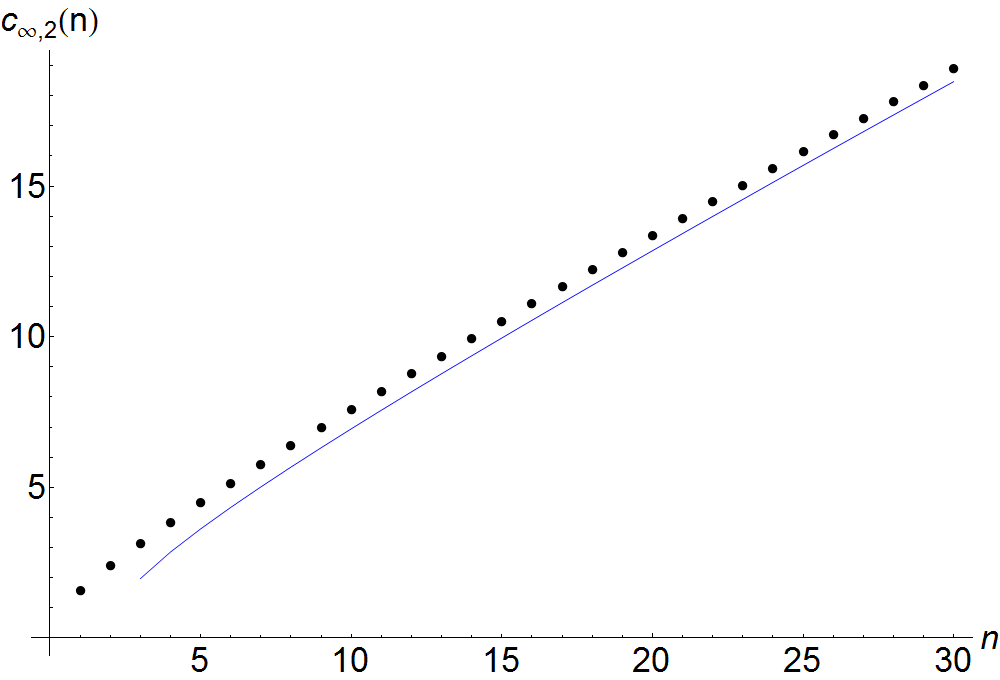}
	\caption{The dots show the dimension dependence of the constant in \eqref{std}, for $\alpha=\infty$, $\beta=2$. The blue line is the asymptotic expansion \eqref{ci2}.
	\label{fig:dimdep}}
\end{figure*}

However, for example, the case $\alpha=\infty$, $\beta=2$ is tractable. For $\alpha=\infty$, $d(\rho^Q)$ is the radius of the smallest ball containing the support of the position distribution. We fix this to be $r=1$, and include other values by scaling symmetry. Then the ``potential'' term in \eqref{radial} becomes zero inside the ball, but diverges outside. Since we are seeking wave functions with finite $d(\rho^P)$, we cannot have a jump at the boundary and must impose zero boundary condition at $r=1$. The bottom eigenvalue $E$ of $P^2$ in \eqref{radial} is then the lowest admissible value of $d(\rho^P)^2$, and either directly from \eqref{std}, or from \eqref{cabfromEab} we find $E=c_{\infty,2}(n)^2$. Clearly, $E$ is lowest for $\lambda=0$, i.e., a purely radial function. At $r=0$ the ground state wave function, written in Cartesian coordinates goes to a constant, so $\phi(r)\sim r^{(n-1)/2}$. This singles out the Bessel function
\begin{equation}\label{bessel}
  \phi(r)\propto \sqrt r\  J_{n/2-1}\bigl(\sqrt E\, r\bigr),
\end{equation}
The scaling $E$ in the argument has then to be chosen so that at $r=1$ we have the first zero $\besselZ1(n/2-1)$ of the Bessel function, which determines the bottom eigenvalue as $E=\besselZ1(n/2-1)^2$. Hence
\begin{equation}\label{ci2}
  c_{\infty,2}(n)=\besselZ1\Bigl(\frac n2-1\Bigr)\approx \frac n2+1.47292\, n^{1/3}-1+{\mathbf o}(1),
\end{equation}
where the asymptotic expansion of $\besselZ1$ is taken from \cite[9.5.14]{abramovich}.
The $n$-dependence of this expression is clearly not as simple as \eqref{22multidim}, although it is asymptotically linear (see Fig~\ref{fig:dimdep}). A direct derivation of this observation will be given below (Sect.~\ref{sec:exStrings}).

\subsection{Number and angle}\label{sec:exNumber}
This case is treated in detail in \cite{KW}. For definiteness, let us think of the discrete variable as position $Q$, and of the angle-valued one as $P$.
Two metrics naturally suggest themselves for either side: For the discrete variable, say two numbers $x,y\in\Ir$ we can look at $\abs{x-y}$, but we may also just be interested in the probability of two numbers coinciding, which is expressed by the discrete metric $1-\delta_{xy}$. For two angles $p,r$ we may either measure angle along the unit circle, i.e., $\abs{p-r+2\pi n}$, with $n$ chosen to minimize this expression,  or the length of the chord through the circle,
$2\abs{\sin(p-r)/2}$. The tradeoff curves are readily computed numerically, but there are few analytic expressions. For example, for the discrete metric on $\Ir$ ($\alpha=1$), the chordal metric for angles ($\beta=2$) we have
\begin{equation}\label{ZxT}
  d(\rho^Q)^2 +d(\rho^P)^2\bigl(4-d(\rho^P)^2\bigr)\geq1.
\end{equation}

\subsection{Qudits: $\Ir_n\times\Ir_n$}\label{sec:exQudit}
In this case the discrete metric is the natural one, especially when one is interested in quantum information coding problems. For the discrete metric $d(x,x')^\alpha=d(x,x')$, so changing the error exponent gives no new information, and we take $\alpha=\beta=1$. In this space of discrete distributions on $n$ points
\begin{equation}\label{Del}
  \Delta=1-1/n,
\end{equation}
is the ``radius'', i.e., the distance from the totally mixed state to a pure state, and hence the largest possible variance. The ``diameter'', i.e., the largest distance between any distributions is $1$, attained at a pair of distinct pure states. It is clear that when position is sharp, momentum has a flat distribution, so the points  $(0,\Delta)$, $(\Delta,0)$ will be in the uncertainty region.

\begin{figure*}[ht]
	\includegraphics[width=0.45\textwidth]{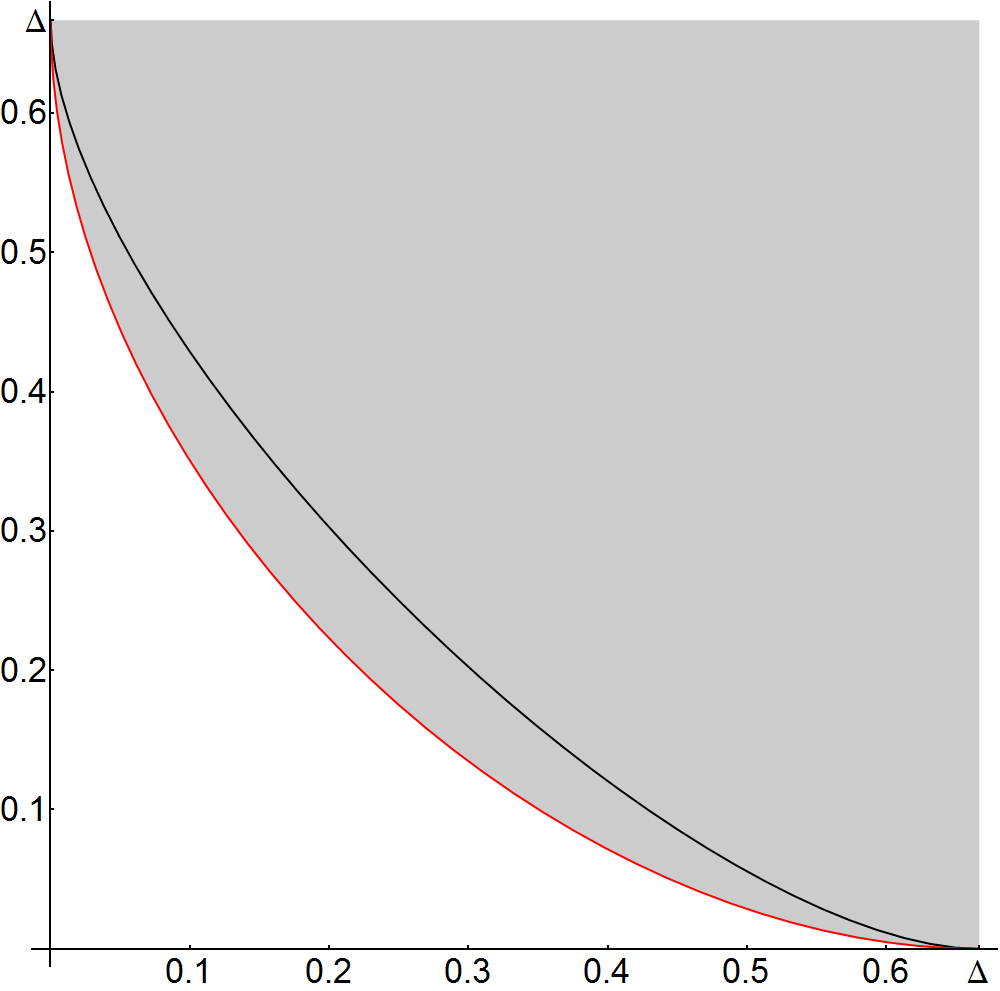}
	\caption{The uncertainty region (shaded) for a qudit as described by \eqref{qudit}, drawn for $n=3$.
     The second ellipse shown inside the region is what is achievable by universal cloning, and measuring on the clones.
	\label{fig:qudit}}
\end{figure*}

Now $d(Q,0)=\idty-\kettbra0$, and  $d(Q,0)=\idty-\kettbra\phi$ with the zero-momentum eigenvector is $\phi=n^{-1/2}\sum_j\ket j$. The ground state of \eqref{linbound} is to be found in the span of $\ket0$ and $\ket\phi$. Hence the pairs of expectations $(\tr\rho d(P,0),\tr\rho d(Q,0))$ are an affine image of a qubit state space, and hence lie on an ellipse, joined with the point $(1,1)$ for states orthogonal to both $\ket0$ and $\ket\phi$. The ellipse fits exactly into the unit square, and also contains the antipodal points $(1-\Delta,1)$ and $(1,1-\Delta)$. This fixes the tradeoff curve (see Fig.~\ref{fig:qudit}).
The resulting uncertainty relation is thus for all $d(\rho^P),d(\rho^Q)\leq \Delta$,
\begin{equation}\label{qudit}
  \Bigl(d(\rho^P)-\Delta\Bigr)^2+\Bigl(d(\rho^P)-\Delta\Bigr)^2+ \Bigl(2-\frac4n\Bigr)d(\rho^P)d(\rho^Q)\leq \Delta^2.
\end{equation}
In this form it is easy to see that if one uncertainty vanishes, the other has to be equal to $\Delta$.

It is interesting to compare this relation, in its version as a measurement uncertainty relation, with a simple ansatz of a joint measurement using the idea of approximate cloning (cf.\ also \cite{Anna}). To this end consider an asymmetric cloner, given by an isometry $V:\HH\to\HH\otimes\HH\otimes\HH$ of the form
\begin{eqnarray}\label{clone}
  V\phi&=&a \phi\otimes\Omega+ b\Omega\otimes\phi, \quad\mbox{with }\\
       1&=& \abs a^2+\abs b^2+ 2\re( \overline a b)/n \nonumber\\
       \Omega&=& n\inv\sum_j\ket{jj}.\nonumber
\end{eqnarray}
Thus $V$ maps systems to three copies of systems, of which the middle one is then traced out as an ancilla. The parameters $a,b$ which of the two output copies is supposed to be the more faithful: When $a=1$ the first copy is just the original system, and the second is completely depolarized, which is reversed for $a=0$. The distinguishing feature of this one-parameter family of cloning maps is the intertwining relation
\begin{equation}\label{itwine}
  VU=(U\otimes\overline U\otimes U)V
\end{equation}
for arbitrary $n$-dimensional unitaries. It implies ``universality'' in the sense that no direction and no basis in Hilbert space is singled out.

When $F$, $E$ are arbitrary full basis projective measurements, we get a joint measurement by
\begin{equation}\label{clonejoint}
  G_{x,y}=V^*(F_x\otimes\idty\otimes E_y)V.
\end{equation}
Its marginals are readily computed to be
\begin{equation}\label{margeclone}
  F'_x:=\sum_y G_{x,y}=(1-\abs b^2) F_x+\frac{\abs b^2}n\,\idty
\end{equation}
Thus $F'$ differs from $F$ by the admixture of state independent noise with a flat distribution and ``probability'' $\abs b^2$. Here the scare quotes indicate that, when $ab<0$, we can have $\abs b^2>1$. The coefficient of $F_x$ may thus negative, but the coefficient of the noise term is always positive.  The largest distance between the output distributions of $F$ and $F'$ is achieved at an eigenstate of $F$. This gives
\begin{equation}\label{distclone}
  d(F,F')=\Delta\,\abs b^2.
\end{equation}

For position and momentum we get a joint measurement, which is also covariant because of \eqref{itwine}. It is therefore generated by a density operator, namely $\rho_F=n V^*(\kettbra0\otimes\idty\otimes\kettbra\phi)V$.
One readily verifies that this is not pure, and hence cannot be optimal. The comparison of the uncertainty pairs generated by cloning and the optimal bound is given in Fig.~\ref{fig:qudit}.

This suggests to relax the intertwining \eqref{itwine} to only phase space shifts. In this way we arrive at a phase space covariant cloning device (not to be confused with a ``phase covariant'' cloner) . Since the phase space structure is the main theme of this paper, we briefly describe how to obtain such maps.
It turns out to be convenient to look not at $V$ but at an operator $\Vh$ with just rearranged matrix elements, which takes $\Cx^n\otimes\Cx^n$ to itself, namely
\begin{equation}\label{Vhat}
  \bra{jkl}V\ket{i}=\bra{jl}\Vh\ket{ki}.
\end{equation}
One then verifies easily that \eqref{itwine} is equivalent to $[U\otimes U,\Vh]=0$, and that \eqref{clonejoint} becomes $G_{x,y}=\tr_1\Vh^*(F_x\otimes E_y)\Vh$, where $\tr_1$ denotes the partial trace over the first factor. Now these relations are only demanded for $U=W(q,p)$, so $\Vh$ lies in the algebra spanned by the Weyl operators commuting with the group of operators $W(q,p)\otimes W(q,p)$. It is hence a linear combination
\begin{equation}\label{Vhexp}
  \Vh=\sum_{q,p} u(q,p) W(q,p)\otimes W(-q,-p),
\end{equation}
where the $u(q,p)$ are suitable complex coefficients. The normalization condition is $\tr_2\Vh^*\Vh=\idty$, which can by guaranteed by adjusting an overall scalar factor, because the left hand side commutes with all Weyl operators, and is hence a multiple of the identity. For the same reason as in the case of the universal cloner, the phase space covariant cloner will give a covariant observable. So in order to explore the possibilities, it suffices to determine the density operators $\rho_F=n G_{0,0}$ obtained by various choices of $u$. Direct computation gives (up to irrelevant constant factors)
\begin{equation}\label{rohF}
  \rho_F=\sum_{q,p,q',p'} u(q',p')\overline{u(q,p)}\ \delta_{qq'}\ \ketbra{p}{p'} =\sum_q \kettbra{\psi_q},
\end{equation}
where $\psi_q=\sum_p\overline{u(q,p)}\ket p$, and the kets $\ket p$ are eigenkets of momentum. Clearly, we can choose $u(q,p)$ so that the $\psi_q$ are the eigenvectors (times the square root of the eigenvalue) of any density operator we choose. It follows that {\it every} covariant observable can be realized by phase space covariant cloning.

\subsection{Qubit strings: $\Ir_2^n\times\Ir_2^n$}\label{sec:exStrings}
In this case ``position'' corresponds to the readout in computational basis, say the product of the $Z$ eigenbases for every qubit, and ``momentum'' is the readout in the product of some conjugate eigenbases, say $X$. As the distance function we take the Hamming distance per qubit:
\begin{equation}\label{Hamming}
  d(x,x')=\frac1n\sum_{i=1}^n\abs{x_i-x_i'}.
\end{equation}
The Hamiltonian is now a many-body operator with non-commuting terms. However, for large $n$ the two terms commute approximately, and the ground state problem is within the scope of  mean-field theory, as laid out in \cite{RW89}. The basic result is that the ground state energy is obtained asymptotically by minimizing instead a classical function on the one-particle state space. We associate with the Hamiltonian $H_{\alpha\beta}(t)$ a ``classical Hamiltonian function''  on the set of one-particle density matrices $\rho_1$, namely
\begin{equation}\label{hclassical}
  \Bigl(h_{\alpha\beta}(t)\Bigr)(\rho_1)= \Bigl(\tr\rho_1 d(P_1,0)\Bigr)^\alpha+t\Bigl(\tr\rho_1 d(Q_1,0)\Bigr)^\beta
\end{equation}
where $P_1,Q_1$ are the one-particle position and momentum. In the limit $n\to\infty$ the ground state energy converges to the minimum of this function. We do not have to compute this minimum explicitly, since we are only interested in the uncertainty region which it outlines. This directly given by the two terms in \eqref{hclassical}, with $\rho_1$ ranging over the one-particle state space. Taking qubits with $X$ and $Z$ measurements now, we parametrize $\rho_1$ by its Bloch sphere coordinates, and find that the boundary curve of the asymptotic uncertainty region $\UR_\infty$ is given by
\begin{equation}\label{qubitstr}
  t\mapsto \Bigl(\bigl({\textstyle\frac12}(1+\cos t)\bigr)^\alpha,\bigl({\textstyle\frac12}(1+\sin t)\bigr)^\beta\Bigr).
\end{equation}

This method works for all systems of a large number of equal copies. We can use it also to get a handle on the dimension dependence in Sect.~\ref{sec:exStandard}. Let $E_n(a,b)$ be the constants in
\eqref{Hab} and \eqref{Eab} with the $n$-dependence made explicit. If we set $a=n^{-\alpha/2}$ and $b=n^{-\beta/2}$, the resulting ground state problem is of mean field type, and we get
\begin{eqnarray}\label{p}
  \lim_n E_n(a,b)&=&\min_{\rho_1}\Bigl\{\bigl(\tr\rho_1 Q_1^2\bigr)^{\alpha/2}+\bigl(\tr\rho_1 P_1^2\bigr)^{\beta/2}\Bigr\}\nonumber\\
   &=&\min_v\bigl(v^{\alpha/2}+(4v)^{-\beta/2}\bigr)
   =(\alpha +\beta )\ 2^{-\frac{\alpha  \beta }{\alpha +\beta }} \alpha ^{-\frac{\alpha }{\alpha +\beta }} \beta ^{-\frac{\beta}{\alpha +\beta }}
\end{eqnarray}
On the other hand, from \eqref{Eab} we get $E_n(a,b)=n^{-\frac{\alpha  \beta }{\alpha +\beta }} E$ and hence from \eqref{cabfromEab} we have
\begin{equation}\label{Enab}
  E_n(a,b)
     = n^{-\frac{\alpha  \beta }{\alpha +\beta }}(\alpha +\beta )\  \alpha ^{-\frac{\alpha }{\alpha +\beta }} \beta ^{-\frac{\beta}{\alpha +\beta }}
       c_{\alpha\beta}(n)^{\frac{\alpha  \beta }{\alpha +\beta }}
\end{equation}
Combining this we get the remarkably simple result that
\begin{equation}\label{limcab}
  \lim_n c_{\alpha\beta}(n)\,\frac2n=1,\quad \mbox{for all}\ \alpha,\beta\geq1.
\end{equation}

\section*{Acknowledgements}
Support by the ERC (grant DQSIM) and the European Commission (project SIQS) and the German BMBF-funded network Q.com-Q is gratefully acknowledged.

%\bibliography{urpslit}

\end{document}